\documentclass[12pt]{article}
\usepackage{amsmath,amssymb}
\usepackage{graphicx}
\def\R{\hbox{{\rm I}\kern-0.2em{\rm R}\kern0.2em}}%definition of reals
\def\R{\hbox{{\rm I}\kern-0.2em{\rm R}\kern0.2em}}%mathematical R for reals
\def\D{\hbox{{\rm I}\kern-0.2em{\rm D}\kern0.2em}}

\def\be{\begin{equation}}
\def\ee{\end{equation}}

\def\({\left(}
\def\){\right)}
\def\[{\left[}
\def\]{\right]}
\def\bc{\begin{center}}
\def\ec{\end{center}}

\begin{document}

{\large \bf Gravitational Wave Sources May Be ``Further" Than We
Think}

\textit{ASGHAR QADIR}$^{\dag}$

$^{\dag}$Centre for Advanced Mathematics and Physics (CAMP)\\
National University of Sciences and Technology\\
Campus of the College of Electrical and Mechanical Engineering\\
Peshawar Road, Rawalpindi, Pakistan

aqadirmath@yahoo.com

{\bf Abstract}. It has been argued that the energy content in time
varying spacetimes can be obtained by using the approximate Lie
symmetries of the geodesics equations in that spacetime. When
applied to cylindrical gravitational waves, it gives a self-damping
of the waves. According to this proposal the energy of the waves go
to zero asymptotically as the radial distance to the two-thirds power.
If true, this would mean that the estimates for the sensitivity of the
detectors for the various sources would have to be revised.

\textit{Key words}: Gravitational waves; SN1987A; Energy;
Approximate Lie symmetries

In the early days of Relativity there were doubts raised about the
reality of gravitational waves as they were solutions of vacuum
field equations. Weber and Wheeler demonstrated that they would
impart momentum to test particles in their path \cite{WW}. This
demonstration was extended to test particles in the path of plane
gravitational waves by Ehlers and Kundt \cite{EK}. A general closed
formula for the momentum imparted to test particles in an arbitrary
spacetime was given by Qadir and Sharif \cite{QS}.

Though it seemed obvious that ``the waves must carry energy", there
was no clear measure available for the energy carried by them. This
is because there is generally no energy conservation in Relativity
and hence no definition of mass or energy. One way to avoid this
problem is by defining a stress-energy ``pseudo-tensor" (see for
example \cite{LL}), but it is observer dependent and not generally
accepted for defining the energy by relativists (see for example the
discussion of the pseudo-tensor in \cite{MTW}). A deeper approach is
to define some measure for the breaking of time translational
symmetry \cite{KA,MR,IR,YJ,TA,BCP,SB} and use that to provide a
measure for the energy density in the spacetime. None of these
attempts showed unambiguous success as they did not lead to any
basic new physical insights or predictions.

Unlike electromagnetic waves, gravitational waves undergo-self
interaction due to the non-linearity of Relativity. This
self-interaction could, in principle cause damping like Landau
damping of electromagnetic waves \cite{Lan}, due to their
interaction with matter, or cause an enhancement such as might be
expected on the basis of the work on colliding gravitational waves
\cite{KP,PS}, where singularities are formed due to the interaction
between two plane gravitational waves.

One approximate symmetry proposal is the use of broken (or
approximate) Lie symmetries \cite{BGI,Ib} of the geodesic equations
\cite{IA}. The breaking is taken to be by terms involving a small
parameter whose powers, higher than some chosen value, can be
neglected. It was found that though with the first order approximate
symmetries no new insight is obtained \cite{KMQ}, for the second
order the timelike symmetry picks up a scaling factor
\cite{IMQ,IMQ1} independent of the strength of the breaking. It may,
thus, provide a genuine measure for the energy content of
gravitational waves. This proposal has been implemented for
cylindrical gravitational waves \cite{IMQ2}. It yields a prediction
that the self-interaction of the waves {\it attenuates} them by a
factor $\sim \sqrt{(c/\rho \omega)}$, where $\rho$ is the radial
distance from the source and $\omega$ is the frequency of the wave.
Note that, $\rho \omega /c=n$ is the number of wave-lengths in the
distance from the source. The longest waves will, thus, contribute
the most in detection.

The purpose of this note is to examine the observational
implications of this attenuation of gravitational waves implicit in
the approximate symmetry proposal and look for predictions that can
be tested.

The obvious implication of the predicted behaviour of energy is that
our expectation of how the energy in the wave decreases with the
distance of the source and the frequency of the wave (and hence the
length of the antenna) will be very seriously overestimated. One can
view this reduction alternatively as an effective increase of the
distance of the source due to the warping of the geometry of the
spacetime by the wave. However, one can not simply take the
reduction factor to be $1/\sqrt{n}$, as that is found for
cylindrical waves. There are no infinite axially symmetric sources
in Nature to emit genuine cylindrical waves. Nor is it absolutely
clear that a spherical, or slightly aspherical, source will give the
same attenuation factor.

An interesting source to consider is SN1987A, which had a 0.1
asphericity in its explosion \cite{WWH}. One could regard the
resulting waves as, in some sense, combined ``spherical" and
cylindrical waves. Crudely, then, one could take a cylindrical
component of 0.1 of the total and assume that it has the predicted
attenuation factor. The axis of the explosion is estimated to make
an angle of $\pi/4 =45^o$ with our line of sight. The energy of
spherical waves decreases as the square of the distance in normal
geometry. It seems reasonable to assume that in the warped geometry
there will be an attenuation $\ge \sqrt{(c/\rho^3 \omega)}$. This
reduction is enormous for SN1987A, which is at a distance of $\sim
51.4~kpc$. For an interferometer with an arm of $5.3~km$ the
reduction factor comes out to be about a million! For the
Hulse-Taylor \cite{HT} binary pulsar it would be about $300,000$ for
the interferometer. Note that due to the frequency dependence of the
attenuation, the long base-line interferometers will be better to
look for gravitational waves. For a bar detector the above
attenuation factor would be up by a factor of about thirty. {\it
There is a clear prediction that the waves from this pulsar will not
be seen by the currently planned detectors}.

Weber claimed to have observed a signal on his {\it non}-cryogenic
bar detector synchronous with the light and neutrino burst from
SN1987A \cite{JW}. Since his (standard) theory of the bar detector
\cite{JW1} leads to energy requirements incompatible with the energy
output of SN1987A ($\sim 10^{53}~ergs$), this claim was generally
regarded as untenable. Indeed, Weber tried to re-vamp his theory of
the detector to provide much greater efficiency of detection, using
a claimed quantum coherence effect \cite{JW2,GP}. While Bassan did
not believe the claim, he admitted to having seen the signal found
by Weber, as he was visiting Weber's laboratory at the time
\cite{MB}. If our proposal is borne out, then Weber's claim of a
sensitivity a billion times greater than that provided by his
original theory would get reduced to an effective thousand --- still
far from sufficient to provide observability.

Of course, there are no exact spherical gravitational wave solutions
available. However, there are Nutku's spherical wave solutions
``with strings attached" \cite{YN}. The approximate Lie symmetry
analysis could be applied to them to obtain the attenuation factor
in that case. This analysis is vital for a more precise test of the
proposal.

A detailed analysis of the sensitivity requirements for various
sources according to their strengths and distances will be presented
subsequently \cite{FQ}.

\section*{Acknowledgments} I am grateful to F. De Paolis for help in
obtaining information regarding SN1987A.

\end{document}